\documentclass[aps,letterpaper,11pt,reprint,floats,showkeys,longbibliography]{revtex4-1}

\usepackage{amsmath}    
\usepackage{graphicx}   
\usepackage{verbatim}   
\usepackage{subfigure}  
\usepackage{hyperref}   

\usepackage[usenames,dvipsnames]{xcolor}

\newlength{\narrowfig}
\newlength{\widefig}

\newcommand{\zhat}{\mathbf{\hat{z}}}

\newcommand{\nhat}{\mathbf{\hat{n}}}

\newcommand{\tDelta}{\widetilde{\Delta}}

\begin{document}

	\title{Effects of multijet coupling on propulsive performance in underwater pulsed jets}

	\date{\today}

	\author{Athanasios G. Athanassiadis}
	\email[Please send correspondence to:\ ]{thanasi@mit.edu}
	\author{Douglas P. Hart}
	\email{dphart@mit.edu}
	\affiliation{Department of Mechanical Engineering, Massachusetts Institute of Technology}
	\begin{abstract}
	Despite the importance of pulsed jets for underwater propulsion, the effect of multiple-jet interactions remains poorly understood. We experimentally investigate how interactions between parallel jets in a pulsed-jet thruster affect the thruster's propulsive performance. Using high-speed fluorescence imaging, we investigate the mutual influence of two pulsed jets under conditions relevant to low-speed maneuvering in a vehicle ($Re\nobreak\approx\nobreak350,$ $L/D\leq2$). Thrust production and propulsive efficiency are evaluated for different nozzle spacings using a new force estimation technique based on the fluorescence data. This analysis reveals that, compared to non-interacting jets, the efficiency and thrust generated by the pair of interacting jets can fall by as much as 10\% when the jets are brought into close proximity. Empirically, the thrust $T$ falls off with the non-dimensional jet spacing $\tDelta$ as $T=T_\infty(1-Co \tDelta^{-6})$ for a thrust coupling coefficient $Co=2.04 \pm 0.11.$ Finally, we predict this dependence of thrust on spacing using a model that relates the thrust and efficiency drop to streamline curvature and vortex induction at the nozzles.
	\end{abstract}
	

	\maketitle

	\section{Introduction}

	A significant limitation for underwater robots is their ability to maneuver precisely during complex sensing and tracking tasks. Next generation vehicles require thrusters that can overcome this problem and efficiently provide precise maneuverability at low speeds. Such maneuverability, in turn, requires thrusters that can deliver specific impulses rapidly and efficiently to the vehicle. In these settings, pulsed jets are increasingly used to augment vehicle maneuverability and improve efficiency at low-speeds. Pulsed jets offer many benefits over traditional propeller propulsion, including more precise impulse delivery, more rapid impulse delivery, and the ability to propel a vehicle using zero mass flux \cite{Mohseni:2006eo}.
	
	Pulsed jets also provide opportunities for efficient individual and swarm propulsion as observed in animals such as squid, jellyfish, siphonophores, and salps. For salps and siphonophores uniquely, individual animals form chains where each member can independently control its jetting behavior. By synchronizing jet strength and timing, colonies of these animals can execute precise maneuvers and can reach high speeds efficiently \cite{Bone:1983uo,Madin:1990iy, Sutherland:2010cy, Costello:2015fn}. Taking inspiration from nature, pulsed jets may prove to be an important technology for the development of scalable marine robotic swarms. However, in order for pulsed jets to be used more widely for underwater vehicles, the implications of jet hydrodynamics on vehicle design must be understood.
		
	 Individual pulsed jets have become the most widely explored form of underwater jet propulsion since it was shown that a pulsed jet will generate more thrust than an equivalent steady jet \cite{Weihs:1977ty, Krueger:2001vo, Krueger:2003cs}. This phenomenon can be explained by breaking the thrust production process into two contributions - one associated with the inertial momentum transfer of a steady jet, and one associated with a nozzle ``overpressure'' generated by the unsteady starting flow as the pulsed jet rolls into a vortex ring. For pulses shorter than the critical vortex formation time, the nozzle over-pressure contributes as much as half of the total impulse generated by the pulsed jet, suggesting it is more beneficial to ``chop'' the flow into short pulses than to eject a steady jet of long duration \cite{Krueger:2001vo, Krueger:2003cs}. Complementary descriptions of thrust production in pulsed jets can leverage concepts of vortex added mass \cite{Weihs:1977ty, Krueger:2001vo, Krueger:2003cs}, or streamline curvature at the nozzle \cite{Krueger:2005dp}.
	 
	To realistically use pulsed jets for marine propulsion, it is necessary to understand how thrust production is affected by different flow conditions. For example, when a pulsed jet is ejected into a background flow parallel to the jet (co-flow), circulation (and hence thrust) production decreases as the speed of the co-flow increases \cite{Krueger:2003fb}. By contrast, if the ambient flow is antiparallel to the jet (counter-flow), the pulsed jet takes longer to separate from the nozzle, increasing the circulation in the  vortex ring and the duration of the over-pressure benefit experienced by the jet \cite{Dabiri:2004cf}. Experiments and simulations have further indicated that circulation production can be controlled through the design, and in some cases real-time manipulation, of the nozzle geometry \cite{Allen:2005kk, Dabiri:2005gu, Rosenfeld:2009ii, Ofarrell:2014ko}.
	 
	For many applications, interactions between multiple jet pulses will strongly affect propulsion when the pulses are closely spaced in time. For these continuously-pulsed jets, thrust production and propulsive efficiency deviate from the single-pulse behavior. Such effects have been characterized as functions of design parameters including system geometry, jet velocity, and pulse frequency \cite{Moslemi:2010ji, Nichols:2012ku, Krieg:2013bm}. Using  experiments on a model vehicle, \citeauthor{Ruiz:2010cd} demonstrated that the vehicle efficiency could be as much as 70\% greater when propelled by pulsed jets than when propelled by steady jets \cite{Ruiz:2010cd}. In follow-up experiments, \citeauthor{Whittlesey:2013fm} used a similar technique to relate such efficiency gains to the wake kinematics \cite{Whittlesey:2013fm}.
	 
	 In addition to temporally-separated jet pulses, the geometric placement of independent jets can affect propulsive performance. It has been shown that multiple pulsed-jet thrusters can be used to improve the control and maneuverability for underwater vehicles \cite{Mohseni:2006eo, Krieg:2008bl}. However in these cases, the jets were spaced far enough apart that there were no hydrodynamic interactions between them. For other designs -- such as small robots with closely-spaced thrusters, or swarms of independent vehicles operating in close proximity -- the hydrodynamic interactions of multiple pulsed jets could affect the thrust and efficiency of the pulsed-jet propulsion. 
	 
	 Despite the importance of multi-jet interactions for underwater propulsion, there are no existing descriptions of these interactions that can inform thruster design. In this paper, we experimentally investigate how thrust production is affected by multi-jet interactions by visually observing the wakes formed by two parallel pulsed jets. We observe that, as nozzle spacing $\Delta$ decreases, the thrust and efficiency fall according to $1-Co\ (\Delta/D)^{-6},$ where $D$ is the nozzle diameter and $Co$ is a dimensionless ``coupling number'' that describes how strongly the two-nozzle coupling affects thrust and efficiency. We explain this dependence with a model based on vortex interactions and geometric constraints in this problem. Our model predicts the observed wake kinematics and reveals the potential for thrust augmentation under certain conditions.
	 	
	\section{Experimental Approach}
	\subsection{Hardware}
	
\begin{figure}[t!]
		\centering
		\includegraphics[width=\narrowfig]{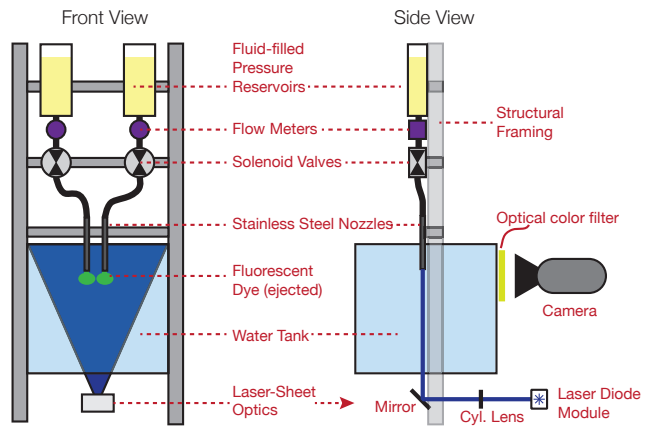}
		\caption{\label	{fig:expt_geom} Experiment schematic depicting all of the system components. The side view reveals the illumination and imaging system, which images the jets from the front.}
	\end{figure}

	In order to understand the evolution of thrust and efficiency in a two-nozzle pulsed jet, we imaged pulsed jet formation at early times, using the motion of fluorescent dye in the wake to estimate hydrodynamic forces. A schematic of the experimental setup is provided in Fig.~\ref{fig:expt_geom}. 

	The jets were created in a (30.4cm)$^3$ cubic tank using two $D=6.35$mm inner-diameter stainless steel nozzles submerged in water. The nozzles were mounted on a linear rail so that nozzle spacing could be varied continuously. Flow through each nozzle was driven by an independent pressure reservoir. For these experiments, the pressure reservoir was hydrostatic, consisting of two open 60mL syringes filled to capacity. Between the reservoir and nozzles, additional hardware measured and controlled the flow in the experiment. Volume flux was measured using a low-inertia, positive-displacement flow meter (FCH-m-POM 97478039; BIO-TECH e.k.), and the flow was controlled by an inline solenoid valve (CNYUXI 2W-025-08).

	When open, the valve constricted fluid flow to a 2mm diameter, and the associated viscous losses in the tubing and valve brought the effective driving pressure to $p_d\approx114$Pa (see Supplemental Content for details). This driving pressure fixed the jet velocity within the range $u_j=48.2-55.2$mm/s for a 0.3s pulse. Using the nozzle diameter as a length scale, the typical jet Reynolds number was $Re=350.$ 
	
	In order to follow the evolution of the developing jets, the reservoirs contained water mixed with fluorescein dye ($5\times 10^{-7}$~M fluorescein sodium salt in water). This way, all of the fluid ejected from the nozzle was marked with a fluorescent tracer, while the fluid in the tank was transparent. The jets were illuminated with a blue (462nm) laser diode module (1.5W optical power; Lasertack LDM-462-1400-C). The laser module emits a 4mm-diameter gaussian beam that was expanded into a 4mm thick laser sheet using a cylindrical lens. This laser sheet was centered on the nozzles to illuminate the central plane of both jets, as depicted in Fig.~\ref{fig:expt_geom}. When exposed to the blue laser sheet, the fluorescein dye emits green light (532nm), which was recorded at 400~frames/second using a high-speed camera (Phantom Miro 320s; Vision Research Inc.). This frame rate corresponds to a temporal resolution of 2.5ms.
			
	The hardware was digitally controlled to synchronized the valves, laser, and camera acquisition. Each experiment lasted for 1~second. First, the camera and laser were triggered, then the valves were opened providing a sudden pressure gradient to initiate the pipe flow. After 0.3s, the valves were closed, and the flow was recorded until 1s had passed from the start of the experiment. Finally, the tank was allowed to settle for 3~minutes before beginning the next experiment. This procedure ensured that no residual vorticity or dye would impact the measurements of consecutive experiments. 
	
	\subsection{Analytical Framework}
			
	\begin{figure}[b!]
		\centering
		\includegraphics[width=\narrowfig]{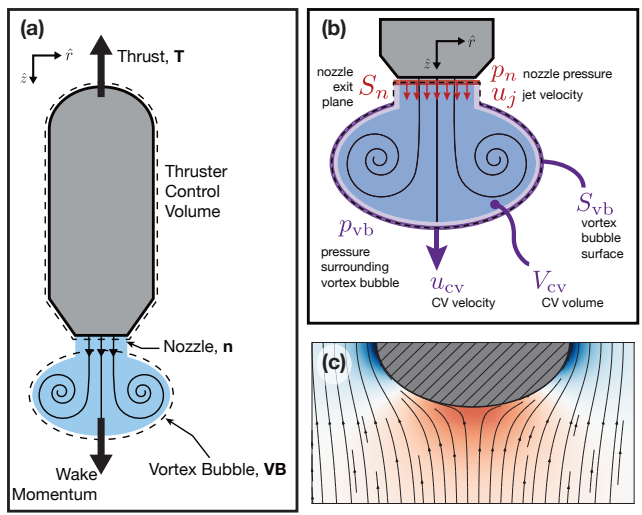}
		\caption{\label{fig:analysis_CV} Estimating thrust based on wake evolution. \textbf{(a)} The control volumes around a model vehicle and its vortical wake. The thrust experienced by the vehicle is balanced by the rate of momentum change in the wake. \textbf{(b)} A detailed control volume $\Omega_{cv}$ around the vortical wake. Two surfaces contribute to the momentum in the $\zhat$ direction: the nozzle exit surface (red), and the time-varying material boundary between the wake and the external fluid (purple). The contributions to the momentum must balance so that the momentum flux through the nozzle exit surface is compensated by a mixture of control volume growth, motion, and pressure along the surface of the nearly-ellipsoidal control volume. \textbf{(c)} Pressure distribution around an ellipsoid moving steadily in the $+\zhat$ (downward) direction. Red coloring indicates positive pressures, blue indicates negative pressures, and white indicates zero pressure. Streamlines of the associated potential flow are drawn in black.}
	\end{figure}
	
	Since the small forces in this experiment ($A_n \rho u_j^2\sim0.1$mN) are difficult to measure directly, we derive an indirect method that allows us to estimate the thrust produced by the jets using a video of the wake kinematics. A similar approach is described by \citet{Ruiz:2010cd}.
	
	The thrust produced by an underwater jet can be calculated from the momentum flux through the nozzle exit plane. Surrounding the nozzle (or thruster) by a control volume (see Fig.~\ref{fig:analysis_CV}a), the forward thrust $T (-\zhat)$ is balanced by the inertial momentum transfer out of the nozzle and the pressure on the nozzle exit plane. Assuming that the jet velocity and pressure are roughly constant along the nozzle exit plane (using the `slug model' of vortex formation), the momentum equation can be integrated around the control surface to yield an equation for the thrust produced by the jet:
	\begin{align}\label{eq:thruster_thrust}
		T &= \int_{S_n} \left[\rho (\mathbf{u}\cdot\zhat)^2 + p_n - p_0\right]\ dA \nonumber \\ 
		  &\approx A_n (\rho u_j^2 + p_n-p_0),
	\end{align}
	
	\noindent where $A_n$ is the nozzle area, $\rho$ is the fluid density, $u_j$ is the fluid velocity exiting the nozzle, $p_n$ is the pressure along the nozzle-exit plane, and $p_0$ is the free-stream pressure. Because a pulsed jet initially rolls up into a vortex ring, for early times the nozzle pressure is not equal to the free-stream pressure, and the `nozzle over-pressure' contributes significantly to thrust production \cite{Krueger:2001vo, Krueger:2003cs}.
	
\begin{figure}[t!]
		\centering
		\includegraphics[width=\narrowfig]{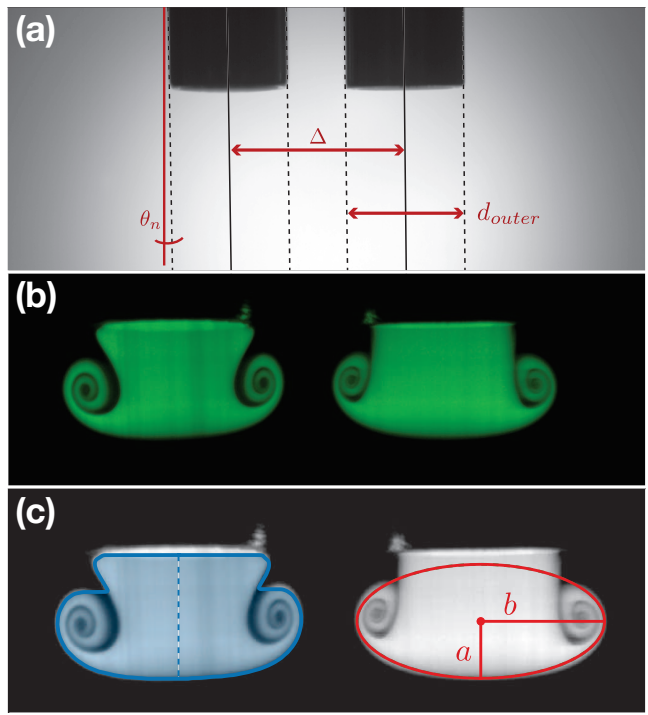}
		\caption{\label	{fig:analysis_vid} Key steps in video processing for thrust estimation. \textbf{(a)} A calibration image of the nozzles is acquired and analyzed to locate nozzle position, size, and camera tilt. These measurements are used to preprocess the images so that they are consistent across imaging runs, and can be analyzed with physical units. \textbf{(b)} Typical image from a two-nozzle experiment ($\tDelta=1.94$). \textbf{(c)} Results of control volume analysis on the experiment images. Left nozzle - the control volume is shaded in blue, with the centerline shown as the vertical dotted line. Right nozzle - the ellipsoid fit to the wake for surface pressure calculations.}
	\end{figure}

	To relate the thrust production to wake kinematics, we consider a second control volume, $\Omega_{cv},$ surrounding the wake. This control volume encloses all of the fluid ejected from the nozzle, as well as any fluid entrained into the vortex structures that form. Given this geometry, the leading region within the control volume is often referred to as a ``vortex bubble'' \cite{Maxworthy:1972ty, Krueger:2001vo, Krueger:2003cs, Olcay:2008fc, Ruiz:2010cd}. The bubble's leading edge defines a material surface that separates the ejected fluid from the ambient fluid initially outside of the nozzle. In experiments, this material surface is easily visualized when dyed fluid is ejected into a clear fluid, as shown in Fig.~\ref{fig:analysis_vid}b. Vortex roll-up during early jet formation causes this material surface to resemble an oblate ellipsoid of revolution, enclosing the forming vortex ring and most of the vorticity in the flow (some of the vorticity diffuses beyond the extent of the material surface \cite{Maxworthy:1972ty, Olcay:2008fc}).
	
	Given this second control volume, the thrust can be calculated from the vertical ($\zhat$) momentum conservation equation:
	\begin{multline}\label{eq:CV_momentum_z}
		0=\underbrace{\frac{d }{d t} \int_{\Omega_{cv}} \rho \mathbf{u}\cdot\zhat\ dV}_\text{term 1 - unsteady flow}
		 + \underbrace{\rho \int_{S_{vb}} (\mathbf{u}\cdot\zhat) \mathbf{u}_{rel}\cdot\nhat\ dA}_\text{term 2 - fluid inertia} \\
		 + \underbrace{\int_{S_{vb}} (p_{vb} + p_0) \nhat\cdot\zhat\ dA}_\text{term 3 - external pressure} 
		 - \underbrace{\int_{S_n} \rho u_j^2 + p_n\  dA}_\text{term 4 - dynamic pressure at nozzle}
	\end{multline}
	Here, the domain $\Omega_{cv}$ is the entire control volume (CV) as highlighted in Fig.~\ref{fig:analysis_CV}b, $S_{vb}$ is the control surface bounding the (roughly) ellipsoidal vortex bubble, $S_n$ is the control surface at the nozzle exit plane, $\nhat$ is the unit normal at each point on the control surface, $\mathbf{u}$ is the local velocity at each point, $\mathbf{u}_{rel}$ is the velocity relative to the control surface, and $p_{vb}$ is the pressure on the surface of the vortex bubble.	To simplify Eq.~\ref{eq:CV_momentum_z}, we leverage several empirical observations and assumptions.
		
	Focusing first on term 1, an analysis provided in the Supplementary Material reduces the unsteady term to $$\frac{d }{d t} \int_{\Omega_{cv}} \rho \mathbf{u}\cdot\zhat\ dV \approx \rho( \dot{u}_{cm} V_{cv} + 2 u_{cm} \dot{V}_{cv} + z_{cm} \ddot{V}_{cv}).$$ Here, $z_{cm}$ is the $\hat{z}$-position of the control volume's center of mass, $u_{cm} = \dot{z}_{cm}$ is the velocity of that position, and $V_{cv}$ is the CV volume.
	
	Second, observe that on most of the control surface $S_{vb},$ the surface evolves with the fluid so that there is no normal flux, and $\mathbf{u}_{rel}\cdot\nhat=0$ on $S_{vb}.$ Over the region of $S_{vb}$ where entrained fluid enters the control volume, it is often assumed that the entrainment is nearly tangential so that even in this region, $\mathbf{u}_{rel}\cdot\nhat=0$  \cite{Ruiz:2010cd, Olcay:2008fc}. Taken together, these observations indicate that the second term of Eq.~\ref{eq:CV_momentum_z} is negligible.
	
	Third, the two pressures in the third term can be separated. Then, the integrated free-stream pressure around this control surface cancels everywhere except for directly below the nozzle. This reduces the integral to: $$ \int_{S_{vb}} p_0 \nhat \cdot \zhat\ dA = \int_{S_n} p_0\ dA, $$ which can be combined with the fourth term of Eq.~\ref{eq:CV_momentum_z}. This final step reveals the modified fourth term as the thrust that is given in Eq.~\ref{eq:thruster_thrust}.
	
	Leveraging these observations, and rearranging terms in Eq.~\ref{eq:CV_momentum_z}, the thrust produced by the jet can be expressed in terms of the dynamics of the jet wake:
		
	\begin{align}\label{eq:thrust_kinematics}
		T & = A_n \rho u_j^2 + A_n p_n \\
		  & \approx
		\underbrace{\vphantom{\int_{A_b}} \rho \dot{u}_{cm} V_{cv}}_\text{I}
		+ \underbrace{\vphantom{\int_{A_b}} 2 \rho u_{cm} \dot{V}_{cv}}_\text{II}
		+ \underbrace{\vphantom{\int_{A_b}} \rho z_{cm} \ddot{V}_{cv}}_\text{III}
		+ \underbrace{\int_{S_{vb}} p_{vb}\, \nhat \cdot \zhat\ dA}_\text{IV}. \nonumber
	\end{align}
		
	By applying momentum conservation to the CV surrounding the jet wake (Eq.~\ref{eq:thrust_kinematics}), the thrust generated by the pulsed jet is broken down into four terms that rely on three measurable quantities. Term I represents the force to instantaneously accelerate the CV the $\zhat$ direction. Terms II and III represents the forces required to add mass to the growing CV, both by injection and entrainment. The first three terms can be measured by tracking the CV volume and center of mass position at each time during its growth. Term IV describes the pressure around the growing vortex bubble that resists its growth. This term can be thought of in terms of the added mass associated with the growing vortex bubble as it pushes all of the external fluid out of the way \cite{Krueger:2001vo, Dabiri:2005ce}. Term IV requires knowledge of the pressure field surrounding the vortex bubble, which is not directly measurable from the fluorescence experiments.
	
	In order to circumvent the need to directly measure pressure on the vortex bubble, we estimate the pressure on the vortex bubble using the common approximation that flow outside the vortex bubble is nearly irrotational. Using this assumption and the observation that the vortex bubble is approximately ellipsoidal (see Fig.~\ref{fig:analysis_vid}c), the pressure on the leading surface of $S_{vb}$ can be equated to the pressure on the surface of an ellipsoid that is translating and expanding unsteadily in a potential flow. Such a pressure field is illustrated with the corresponding streamlines in Fig.~\ref{fig:analysis_CV}c. As expected, the pressure field resembles that around a translating sphere, deformed to match the ellipsoidal geometry.
		
	This potential flow model reduces the pressure term (Eq.~\ref{eq:thrust_kinematics}-III) to a function of vortex bubble motion and geometry, which can be tracked as shown in Fig.~\ref{fig:analysis_vid}c. With this simplification, the total thrust can be estimated entirely based on the motion and growth of the wake. 
	
	\subsection{Video Processing for Thrust Estimation}

	To estimate forces from the video data, images are acquired and processed to identify the time-dependent control volume surrounding the wake (see Fig.~\ref{fig:analysis_vid}c, left). First, a calibration image (Fig.~\ref{fig:analysis_vid}a) is acquired and used to determine camera orientation and nozzle positions as shown in the figure. These measurements are used to rotate the raw video data to a standard orientation, and to crop the video into two separate frames - one of each nozzle. Next, preprocessed images are filtered and segmented to identify the CV. In addition to the CV, a bounding ellipsoid is identified for each wake (Fig.~\ref{fig:analysis_vid}c). Finally, the (axisymmetric) volume and center of mass position are numerically calculated and used to calculate the different terms in Eq.~\ref{eq:thrust_kinematics}. While the first three terms can be calculated directly from the CV size and motion, the pressure term requires further computation.
	
	The pressure field surrounding the ellipsoid is calculated by using the unsteady Bernoulli equation in conjunction with the velocity potential around a translating ellipsoid. The calculated pressure field is then numerically integrated around the leading edge of the vortex bubble directly below the nozzle (in 3D, assuming axisymmetry) where the pressure field and $\hat{z}$-component of the surface normal are greatest. A more detailed description of the pressure analysis technique can be found in the Supplemental Content and in \citep{Athanassiadis:2016vu}. As will be discussed in Section~\ref{sec:results:two_noz}, the pressure term ultimately contributes less than 10\% of the total thrust at early times, meaning the unsteady flow terms dominate the early-time wake dynamics.
	
	\section{Results}
	We track the evolution of jet wakes for individual jets \nobreak{($\Delta\to\infty$)}, and for select nozzle spacings increasing from {$\Delta=1.5 D.$} Because $\tDelta = \Delta/D$ arises as the relevant dimensionless group the new geometry introduces, we use $\tDelta$ instead of $\Delta$. For each value of $\tDelta$ tested, five experiments were performed. Videos from the experiments are available in the Supplemental Content.

\begin{figure}[t]
		\centering
		\includegraphics[width=\narrowfig]{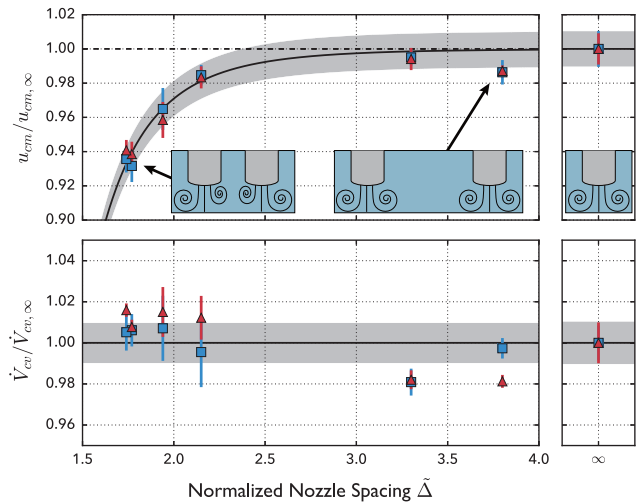}
		\caption{\label	{fig:evolution} Average CV velocity and volume during a 0.3s pulsed jet for different nozzle spacings $\tDelta.$ These results are normalized by the single nozzle values for each nozzle (red triangles represent the right nozzle, blue squares represent the left nozzle). Pulsed-jet interactions cause the wake velocity to drop according to the form $1-C_u\tDelta^{-6}$ (fit shown in figure as solid black line). Error bars on the fit reflect the standard deviation of experimental values $u_{cm,\infty}.$}
	\end{figure}

	For each two-nozzle experiment, the two jet wakes are analyzed independently, providing two measurements of thrust generation for each experiment. Because the nozzles were not perfectly identical, their wakes varied slightly, producing differences in the measurements. To account for this variation in our plots, we eliminate the effects of nozzle fabrication by normalizing each measurement by the average value measured at $\tDelta\to\infty.$ In Figs.~\ref{fig:evolution} and \ref{fig:thrust}, the results from nozzle 1 (left nozzle) are indicated by blue squares, and those from nozzle 2 (right nozzle) are indicated by red triangles. For all experiments, the wake is analyzed between $t=0.08$s and $t=0.30$s. Because the initial startup flow is slow, analysis of the frames before 0.08s did not produce consistent results. The time $t=0$ corresponds to the moment when the solenoid valves are opened and the jets begin to develop.
	
	\subsection{Single Nozzle Dynamics}
	To validate the analysis technique described above, we calculate the scale of the force estimated from the wake of individual pulsed jets, corresponding to the limit $\tDelta\to\infty.$ In these experiments, our analysis estimates a time-averaged thrust of $\overline{T}_\infty=0.10 \pm 0.01$mN, which closely matches the expected thrust scale $T_{exp} \sim \rho u_j^2 A_n = 0.07$mN. The measured average thrust is slightly higher than this predicted scale because of the positive nozzle over-pressure during jet formation.
	
	The wake kinematics for a single jet provide insight into the physical mechanisms driving thrust production in the pulsed jet. First, both the CV center of mass $z_{cm}$ and volume $V_{cv}$ grow nearly linearly with time throughout the experiments, so that the higher derivatives $\dot{u}_{cm}=0$ and $\ddot{V}_{cm}=0.$ This experimental observation has important consequences for the thrust calculation process, reducing Eq.~\ref{eq:thrust_kinematics} to just two terms (II and IV). Additionally, the pressure  term (term IV)  is 10 times smaller than the unsteady term (term II), indicating that flow unsteadiness within the wake is the primary source of thrust during early times. These observations persist in the two-nozzle experiments, supporting the same conclusions across all the experiments (see Supplemental Content for supporting plots).
	
	\subsection{Two-Nozzle Dynamics}\label{sec:results:two_noz}
	As two nozzles are brought into close proximity, their interactions lead to changes in their kinematics as shown in Fig.~\ref{fig:evolution}. In these plots, the multi-jet interactions manifest as a significant change in the wake velocity, but not in the wake volume. Empirically, the wake velocity drops from the single-nozzle value $u_{cm,\infty}$ according to $u_{cm}(\tDelta) / u_{cm,\infty} =  1 - C_u \tDelta^{-6},$ where $C_u=1.86\pm0.06$ is a dimensionless constant that reflects how strongly the two pulsed jets interact. The solid line in the plot represents the (one-parameter) fit of this form to the data, and the shaded region represents the standard deviation of the experimental values $u_{cm,\infty}$.
	
\begin{figure}[t]
		\centering
		\includegraphics[width=\narrowfig]{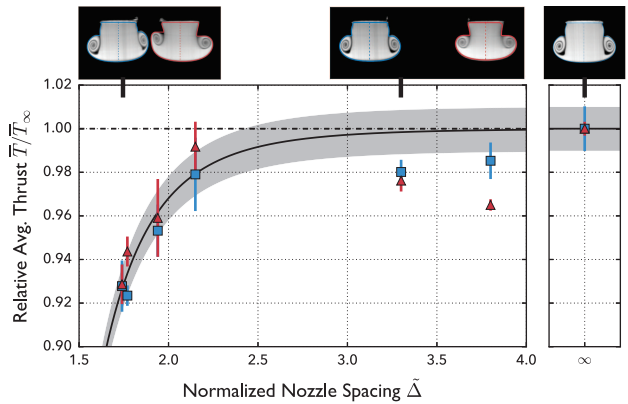}
		\caption{\label	{fig:thrust} Time-average thrust produced during a 0.3s pulsed jet for different nozzle spacings $\tDelta.$ Pulsed-jet interactions reduce thrust production for very close spacing, according to $T/T_\infty = 1-Co\tDelta^{-6}$ (fit shown in figure as solid black line). Here, $Co=2.04$ is a dimensionless `coupling coefficient' that describes how strongly jet interactions affect thrust production. As discussed in the text, this plot also identically reflects the behavior of propulsive efficiency $\overline{\eta}(\tDelta)/\overline{\eta}_\infty.$\vspace{-2em}}
	\end{figure}

	When the wake kinematics are used to calculate the thrust as described in Eq.~\ref{eq:thrust_kinematics}, the thrust generated by each nozzle is observed to follow the same form as the wake velocity, as shown in Fig.~\ref{fig:thrust}. As with the single-nozzle experiments, the thrust from the pressure term (Eq.~\ref{eq:thrust_kinematics}-IV) is 10 times smaller than the thrust associated with unsteady motion in the CV, indicating that the flow unsteadiness within the CV dominates thrust production.
	
	As the two jets are brought together, the total thrust is reduced by nearly 10\%. Empirically, the dependence of thrust on nozzle spacing can be described by the same equation as the wake velocity dependence, with a different coupling coefficient. In this case
	\begin{equation}\label{eq:EmpericalThrust}
		T(\tDelta) / T_\infty  = 1 - Co \tDelta^{-6},
	\end{equation}
	for a thrust coupling coefficient $Co=2.04 \pm 0.11.$
	
	Given the experimental data in Fig.~\ref{fig:thrust}, this model over-predicts the two points where $\tDelta>3.0.$ However, this inconsistency can be seen as arising from the lower volume flow rate in the experiments at those values of $\tDelta$ (see Fig.~\ref{fig:evolution}), and not from the emergence of an unexplained physical phenomenon at those length scales. This explanation is supported by the assumption that the thrust $T$ should smoothly and monotonically asymptote to the single-nozzle value $T_\infty.$
						
	\section{Discussion}
	
	Our results indicate that the interactions between two parallel pulsed jets are a very local phenomenon. When nozzles are separated beyond a few nozzle diameters, wake interactions do not measurably affect thrust production. However, when two simultaneously pulsed jets are brought into close proximity, their wakes destructively interfere with each other, causing the thrust produced by each jet to drop by as much as 10\%. In this section, we describe a physical mechanism that accounts for this observed thrust drop with nozzle spacing.
	
	\subsection{Physical Mechanism for Interactions}
		
	To build an intuition for the physics in the two-jet system, it is helpful to recall some of the observations about the kinematics of the control volumes that led to the thrust estimates above. Observation of the wake growth revealed that $z_{cm}$ and $V_{cv}$ both grow linearly with time at early times. This behavior indicates that $u_{cm}$ and $\dot{V}_{cv}$ are the only nonzero derivatives, so that the unsteady terms Eq.~\ref{eq:thrust_kinematics}I and III are zero. Further, observing that the surface pressure (term IV) is 10 times smaller than the unsteady forces within the wake (term II), we can write that to leading order, the wake dynamics should behave according to $T\approx 2\rho u_{cm} \dot{V}_{cv}.$ Since the driving jet velocity, $u_j$ and the wake volume growth rate $\dot{V}_{cv}$ do not change significantly as the nozzles are brought together, the interactions between the nozzles do not affect entrainment in the wakes. Therefore, the scaling behavior of the thrust production is set by the average velocity within the wake, which we have shown is characterized by the wake center of mass velocity $u_{cm}.$ As the nozzles are brought closer together, this velocity decreases, so the thrust should decrease with an identical form. Physically, this picture is consistent with the hypothesis that vortex induction drives the two-jet interaction: if two separate, coplanar vortex rings are established in an infinite fluid, the circulation in each would establish a counter-flow that lowers the average velocity of the other.
	
	Guided by this intuition, we more rigorously derive a mechanism for the thrust reduction observed in our experiments.
		
	\subsubsection{Scaling as $\tDelta^{-6}$}
	To predict the thrust scaling, we recall the result from Eq.~\ref{eq:thrust_kinematics} that for a single nozzle $$T \sim A_n (\rho u_j^2 + p_n-p_0). $$ Since $\rho,$ $u_j,$ and $p_0$ remain constant as $\tDelta$ is varied, the thrust should scale as the nozzle pressure $T(\tDelta) \sim p_{n,\infty} (\tDelta).$ When a second pulsed jet is introduced into the problem, the nozzle pressure should be modified by a new pressure scale introduced in the problem. In this case, one new pressure scale introduced is that of the vortex ring formed at the other nozzle. Because the velocity induced by a toroidal vortex ring scales as $u_{ind}(r)\sim r^{-3}$ (derivation in supplemental content), the induced pressure should scale as $p_{ind}(r)\sim\rho u_{ind}^2\sim r^{-6}$. Therefore, we can write the thrust produced by a single pulsed jet when a second is located a distance $\Delta$ away:
	
	\begin{align*}	
	T(\Delta) &= A_n (\rho u_j^2 + p_n(\Delta)) \\
	&= A_n (\rho u_j^2 + p_{n,\infty} \pm p_{ind}(\Delta)) \\
	&= T_\infty \pm C \Delta^{-6}.
	\end{align*}

	Reorganizing and and non-dimensionalizing the right hand side of this equation recovers the functional form observed in our experiments (Eq.~\ref{eq:EmpericalThrust}). From here, the sign of the $T$ vs $\Delta$ relationship remains to be determined, corresponding to an expectation for thrust reduction or augmentation.
	
	\subsubsection{Geometric Argument for Thrust Reduction}
	
	In order to predict whether we should expect thrust to increase or decrease from the two-jet interaction, we consider the analysis by \citeauthor{Krueger:2005dj}, which relates the nozzle pressure to the curvature of streamlines at the nozzle. A higher streamline curvature at the nozzle corresponds to a higher average nozzle pressure, according to the equation \cite{Krueger:2005dj}: 
	\begin{equation}\label{eq:streamlines}
		p_n \sim \rho \int_0^{D/2} u_z \frac{\partial u_r}{\partial z}\Big|_{z=0} dr
	\end{equation}
	In the case of two simultaneous pulsed jets, we expect that the interaction between the two forming vortex wakes reduces the streamline curvature at the nozzle exit planes. To see this, consider that the presence of a second nozzle introduces a symmetry plane between the nozzles, across which there can be no volume flux. Accordingly, the flow must adjust to satisfy the zero-flux condition at the symmetry plane, which will restrict the radial growth of the wake. Since the radial flow $u_r$ is restricted, the term $\partial u_r/ \partial z$ should be reduced. This effect can be interpreted as forcing the streamlines coming out of the nozzle to straighten, thereby lowering the nozzle over-pressure $p_n$ according to Eq.~\ref{eq:streamlines}. Therefore, the thrust produced by interacting jets should be lower than that of a single pulsed jet.		
	
	\subsubsection{Additional comments on this model}
	
	This physical mechanism for thrust reduction is consistent with observations in previous experiments on single-nozzle pulsed jets. When a pulsed jet is exposed to ambient co- or counter-flow, the production of circulation (a proxy for nozzle over-pressure) is decreased or increased (respectively) as described by the results of Krueger, Dabiri, and Gharib \cite{ Krueger:2003fb,Dabiri:2004cf}. 
	
	In the case of ambient co-flow (flow parallel to jet), our model would predict that additional stream-wise flow should increase the axial extent of the wake ($a$ decreasing radial growth ($b$) as a result of continuity. Based on these assumptions, our geometric model predicts that the nozzle pressure should decrease, thereby lowering the production of thrust and circulation as is reported by \citet{Krueger:2003fb}. Conversely, in the case of ambient counter-flow (flow antiparallel to the jet), the predictions would reverse, suggesting that the wake should be `flattened' by the counter-flow (i.e. $a$ decreases and $b$ increases by continuity). In that case, the geometric model would predict a higher nozzle pressure, thrust and circulation production, which is reflected in the data from \citet{Dabiri:2004cf}.
	
	\subsection{Efficiency Considerations}
	
	The impact of pulsed-jet interactions on propulsive efficiency can be estimated by considering a conceptual thruster such as the one illustrated in Fig.~\ref{fig:analysis_CV}a. To produce thrust, the jet is driven by a pressure $p_d,$ which ejects fluid at a rate $\dot{Q}.$ Then the power put into generating the thrust is $\dot{E}= p_d \dot{Q}.$ 

	As a measure of useful work, we consider that the goal of these thrusters is maneuverability - rapid bursts of (well-defined) thrust for short periods of time. So a measure of useful work is the thrust produced $T,$ multiplied by the speed at which it can be delivered $u_j$, so that the efficiency of the thruster is given by: $$ \eta = \frac{T u_j}{p_d \dot{Q}} = \frac{T}{p_d A_n}. $$
	
	Since the nozzle area $A_n$ and driving pressure $p_d$ are independent of $\Delta,$ the efficiency of the thruster should scale as the thrust $T$ does. As a result, compared to the efficiency of a single jet $\eta_\infty,$ the efficiency of two jets separated by a distance $\tDelta$ will be given by:
	\begin{equation}\label{eq:efficiency_thrust_relation}
		\frac{\eta (\tDelta)}{\eta_\infty} = \frac{T (\tDelta)}{T_\infty} = 1 - Co \tDelta^{-6}.
	\end{equation}

	\noindent Based on this result, Fig.~\ref{fig:thrust} indicates not only how thrust varies with nozzle spacing, but also how the efficiency should vary with nozzle spacing as well.
	
	\section{Conclusions}
	
	The importance of multiple pulsed jets for underwater propulsion has led us to investigate the effects of multi-jet interactions on thrust production and efficiency for simultaneous pulsed jets. We have developed a control volume approach to estimate thrust production from videos of interacting pulsed jets. This analysis has shown that for intermediate Reynolds number and low stroke length, thrust production is dominated by unsteadiness within the growing wake. When two jets are brought into close proximity, vortex interactions between the jets force the streamlines to straighten, causing the thrust to drop as much as 10\%. However, this effect is highly localized, depending on the nozzle spacing as $\tDelta^{-6},$ so that in practice, a designer should not be concerned with interactions between simultaneous jets separated by more than 2.5 diameters.
		
	Given the data and scaling arguments presented above, the problem remains to determine what sets the coupling coefficient $Co.$ Are there system configurations that can allow $Co<0$? Such behavior would provide a means to increase the pulsed jet's thrust and efficiency through clever system design or control.

	One approach to thrust augmentation is suggested by the role of streamline curvature. If instead of being ejected simultaneously, jets are pulsed so that the second jet is ejected as a stopping vortex forms in the first jet, the close proximity of the negative vorticity from the first jet may \emph{help} to curve the streamlines exiting the second nozzle, thereby augmenting the nozzle over-pressure. In this way, well-timed pulses could exploit  the stopping vorticity in each others' wakes to roll-up more efficiently and produce more thrust. This behavior is reminiscent of how jellyfish exploit stopping vortices to move more efficiently \cite{Gemmell:2013ve}.

	The analysis presented here is not limited to two nozzles. When considering how the nozzle interactions scale to systems with more nozzles (such as a multi-jet vehicle, or a swarm of small vehicles), the geometric and pressure scalings should still apply in a multi-nozzle system. Additional jets will further straighten the flow, and provide additional pressure scales that can be added linearly to the nozzle over-pressure. The most significant difference will be observed because of the 3D nature of the wake development. While the pressure scales can be added linearly, and treated pairwise, the streamlines will develop based on a more complex three-dimensional flow field around the nozzle, and this behavior cannot be predicted by the current analysis.
	
	While we have focused here solely on single-pulsed jets for applications to precision impulse delivery, the interactions between multiple jets can have interesting consequences for continuous pulsed jets. For instance, when our results are related to the analysis and observations by \citeauthor{Ruiz:2010cd} and \citeauthor{Whittlesey:2013fm}, it is reasonable that the changes in wake geometry we observe will have a significant effect on propulsive efficiency in a continuous pulsed jet. Such open questions present exciting opportunities for further research on the interactions of multiple pulsed jets in different operating regimes.
			
	\begin{acknowledgments}
	AGA thanks A. Helal, J. Alvarado, A. Nasto, S. Sroka, and C. Wagner for valuable discussions and feedback. This work was supported by the Lincoln Laboratory and the Office of Naval Research through Award No. 7000308296.		 
	\end{acknowledgments}

	\bibliography{bibliography}

\end{document}


\title{Supplemental content: Effects of multi-jet coupling on propulsive performance in underwater pulsed jets}
	\date{\today}
	\author{Athanasios G. Athanassiadis}
	\author{Douglas P. Hart}
	\maketitle
	
	\vspace{-0.01em}
	\section{Additional Experiment Details}
	\subsection{Nozzles Fabrication}
	The nozzle bores were machined out to the final diameter from stock with a smaller bore. The outer diameter of the nozzles was 9.52mm, and the faces of the nozzles were machined smooth and deburred, but were not tapered to a fine angle or sharp edge as in other experiments \cite{Krueger:2003cs, Krueger:2005dj, Maxworthy:1972ty}. The reservoir was connected to the nozzle with 6.35mm Tygon tubing. Differences in the machining of the two nozzles caused variations in the wakes generated by each nozzle. These differences caused the measurements of wake velocity and thrust to vary slightly between the two nozzles. As described in the text, the effects are accounted for by normalizing the measurements by the single-jet quantity for each jet.

	\subsection{Nozzle Positioning and Driving Pressure}		
	In our experiment, the separation between the tank walls and the nozzle was 120mm, and the nozzle exit plane was submerged 40mm below the free surface. With this configuration, edge-effects are considered negligible during the vortex formation. 
	
	During a typical experiment, the height of the fluid in the reservoirs changed by less than 0.5mm, allowing the change in pressure head from surface motion to be ignored relative to the total reservoir height. The reservoirs were elevated $H=570$mm above the water surface for an expected pressure head of $p_h=\rho g H\approx5500$Pa. However, as described in the Supplemental Materials, losses between the reservoir and the nozzle reduced the effective pressure head to $p_{h,eff}\approx114$Pa.
	
	Given the geometry, the hydrostatic pressure at the nozzle exit is $p_{n,g} = \rho g (40\text{mm})=392\text{Pa}.$     Because of the pressure loss through the tubing and valve, the effective driving pressure was estimated using a simplified momentum equation for unsteady pipe flow:
	$$p_d - p_{n,g} \approx \frac{\rho u_j H}{t} = 114\text{Pa}.$$	
	
	\subsection{Jet Velocity}
	
\begin{figure*}
		\centering
		\includegraphics[width=0.4\textwidth]{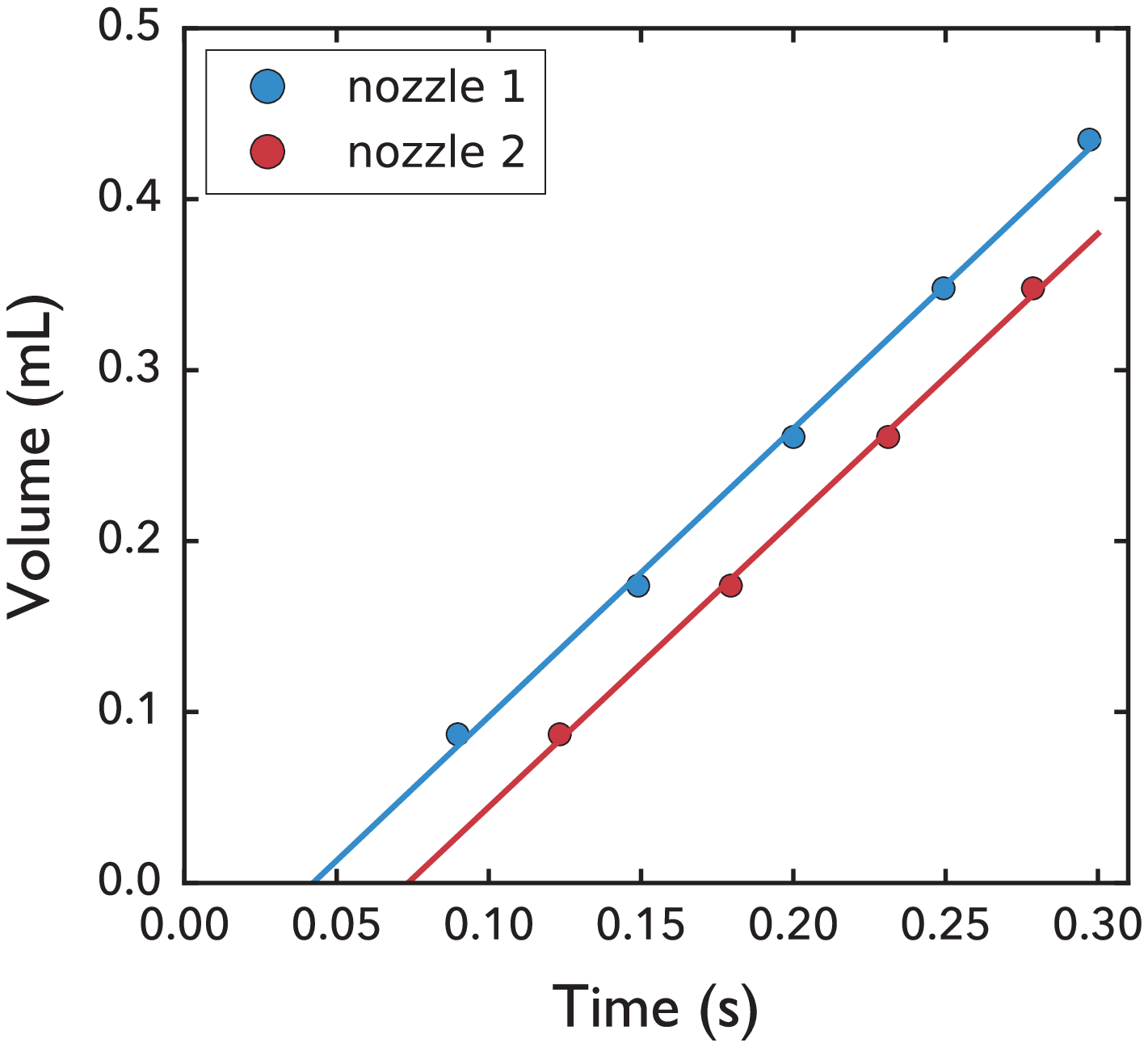}
		\caption{\label{fig:supp:FlowMeter} Volume flux through the nozzle as measured by the flow meters attached to each reservoir. The volume flow through the nozzles is approximately linear in time, so to calculate the volume flow rate, the data for each nozzle are fit to a line. Here, the volume flow rate through each nozzle is $\dot{Q}=1.68\pm0.04$mL/s. The vertical offset of the two traces reflects different initial readings in the flow meters, which are irrelevant for the flow rate calculations.}
	\end{figure*}

	To support our claim in the text that the jet velocities were nearly linear over the course of our experiment, Fig.~\ref{fig:supp:FlowMeter}a shows a typical trace of the flow-meter output for two single-nozzle experiments. As the nozzle spacing was varied, we observed no meaningful change in the jet velocity for either nozzle. 
	
	\subsection{Optical Filtering}

	To prevent scattered laser light from affecting the optical signal, a 495nm optical high-pass filter (Thorlabs FGL495S) was placed in front of the camera.
	
	\section{Analysis of Unsteady Thrust Term}
	
	Our dye-based analysis requires a way to calculate the unsteady term in the $\zhat$ momentum equation, which is reproduced here for convenience: $$ \frac{d}{dt} \int_{\Omega_{cv}} \rho \mathbf{u}\cdot\zhat\ dV.$$ As written, this term requires knowledge of the velocity field $\mathbf{u}$ everywhere within the control volume, which is not accessible through our dye measurements. 
	
	In order to measure flow unsteadiness from dye motion, we rewrite the integral in a Lagrangian formulation so that the local velocity $u_z$ is the time derivative of the $\zhat$ position of a particle within the CV. Then, we can write:
	
	\begin{align*}
	\int_{\Omega_{cv}(t)} u_z\, dV &= \int_{\Omega_{cv}(t)} \frac{\partial Z(\mathbf{r}, t)}{\partial t}\,dV \\
	&=\frac{d}{dt}\left[\int_{\Omega_{cv}(t)} Z\ dV\right] - \oint_{S_{cv}(t)} Z\ (\mathbf{u} \cdot\mathbf{\hat{n}})\ dA \\
	&= \frac{d}{dt} \left[ z_{cm} V_{cv} \right]
	\end{align*}

	The jump from the first to second line is made using the Reynolds Transport Theorem. We expect the surface integral  to be zero because the control surface tracks the dye motion (this assumption validated below). The final step is completed by observing that the integral in the first term is related to the $\zhat$ position of the center of mass $$z_{cm} = \frac{1}{V_{cv}}\int_{\Omega_{cv}} Z dV.$$ Inserting this derived quantity into the unsteady term above yields the result claimed in the main text:
	\begin{equation}\label{eq:supp:unsteadiness}
	\frac{d }{d t} \int_{\Omega_{cv}} \rho \mathbf{u}\cdot\zhat\ dV = \frac{d^2}{dt^2}[z_{cm} V_{cv}] = \dot{u}_{cm} V_{cv} + 2 u_{cm} \dot{V}_{cv} + z_{cm} \ddot{V}_{cv}.
	\end{equation}
	
\begin{figure*}[b]
		\centering
		\includegraphics[width=0.75\textwidth]{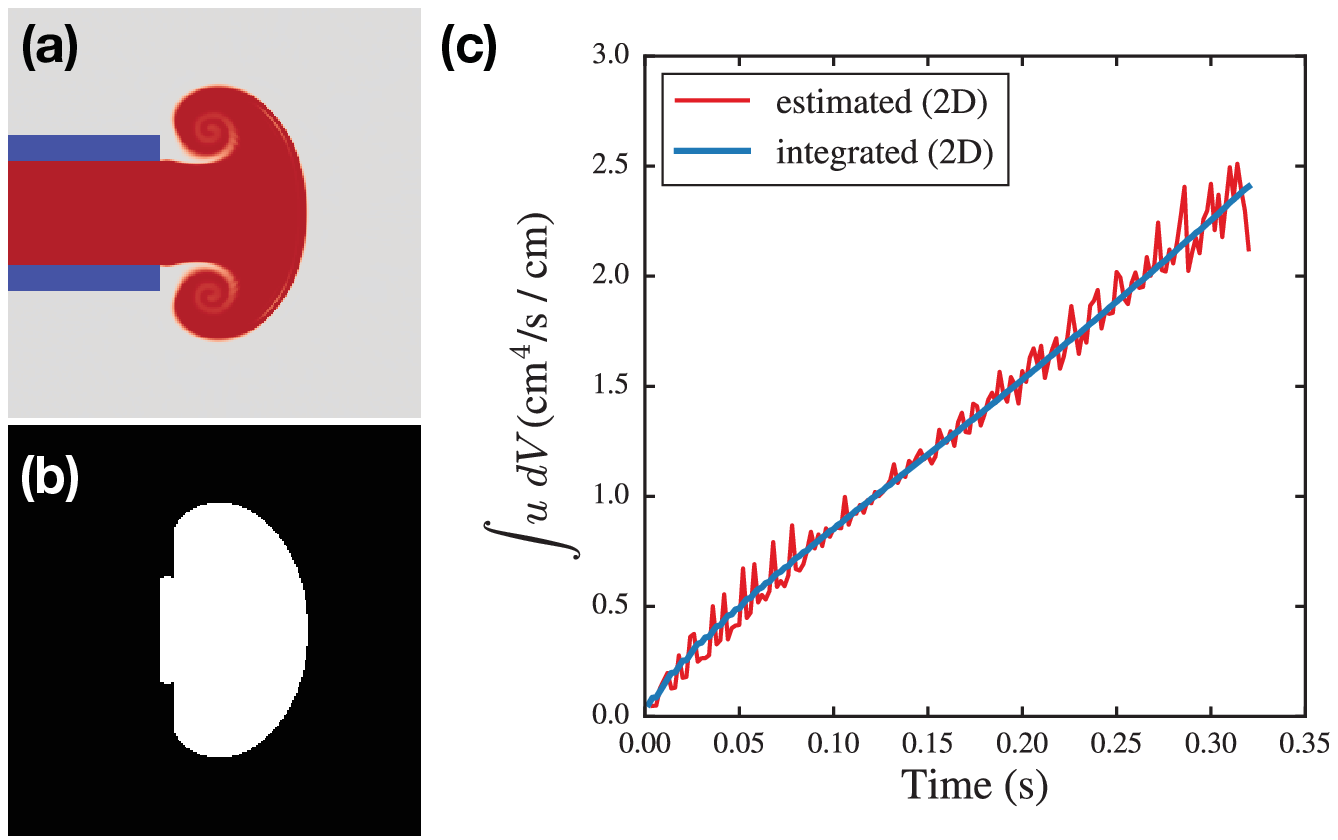}
		\caption{\label{fig:supp:sim} Results of a 2D simulation used to verify our simplification of the unsteady integral. \textbf{(a)} A passive tracer is convected in the simulated flow to produce an image that can be tracked like the experiments.  \textbf{(b)} Using the same analysis technique as for the experiments, the control volume is defined based on the dye location. \textbf{(c)} Comparison between direct integration of the velocity field within the CV and our estimate for the integral \eqref{eq:supp:unsteadiness}. The two calculation techniques overlap exactly within numerical fluctuations that arise from the discrete derivatives required by our simplification.}
	\end{figure*}

	We validated this analysis using 2D finite-volume simulations of a jet exiting a nozzle. By inserting a passive tracer into the nozzle before initiating flow, we can calculate a control volume equivalent to the one used in the experiments. This control volume was then used to calculate the unsteady term by directly integrating the flow field, as well as by using the CV volume and center of mass position. The two independent approaches produced identical results as shown in Fig.~\ref{fig:supp:sim}.
	
	\newpage
	\section{Video Processing}

	\subsection{Segmentation and Region Identification}
	Once the images are aligned and preprocessed, they are filtered and segmented using a local Otsu threshold \cite{Otsu:1975wu, scikit-image} to identify the regions of dyed fluid. The segmented regions are finally filtered with a horizontal hole-filling algorithm to produce a control volume such as the one shown in Fig.~3c in the main text.	
	
	The ellipsoid is identified such that the ellipsoid major radius $b$ is half of the maximum wake width. The minor axis $a$ lies along the nozzle centerline from the major axis to the leading edge of the wake. Since the wake is assumed to be axisymmetric, these two axes uniquely define the bounding ellipsoid as shown on the right of Fig.~3c in the main text.
	
	\section{Leading Edge Pressure Estimation Technique}
	
	This approach was inspired by the work of \citeauthor{Munk:1924tw} on potential flow around ellipsoids for airship design \cite{Munk:1924tw}. In order to compute the last term of the thrust equation (Eq.~3 in the main text), we estimate the pressure on the leading edge of the control surface by evaluate the velocity potential $\phi$ around a translating oblate ellipsoid of revolution. Using the Bernoulli equation:

	\begin{equation}\label{eq:bernoulli}
		 p_{vb} - p_0 = \frac{\rho}{2}(\mathbf{\nabla \phi})^2\Big|_{S_{vb}} + \rho \frac{\partial \phi}{\partial t}\Big|_{S_{vb}}.
	\end{equation}

	We now derive a technique to analytically evaluate the velocity potential in terms of ellipsoid geometry (principal axes $a$ along $\zhat$ and $b$ along $\rhat$) and velocity ($w \zhat$), which are measured as a function of time based on dye motion.
	
	As given by \citeauthor{Lamb:1945tl}, the velocity potential for a translating ellipsoid with principal radii $a$ in the $\zhat$ direction, $b$ in the $\yhat$ direction, and $c$ in the $\xhat$ direction, moving at velocity $w$ in the $+\zhat$-direction is most efficiently described in ellipsoidal coordinates \cite{Lamb:1945tl}:
	
	\begin{equation}\label{eq:ellipsoid_potential_lamb_1}
		\phi(x,y,z,t) = - z\ \cdot\ 
		\underbrace{\frac{w}{2-\alpha_0}}_{C(t)}\ \cdot\ 
		\underbrace{a b c \int_\lambda^\infty \frac{d\lambda'}{(a^2 + \lambda')^{3/2}(b^2+\lambda')^{1/2}(c^2+\lambda')^{1/2}}}_{\alpha(\lambda, a,b,c; t)}.
	\end{equation}
	
	In Eq.~\ref{eq:ellipsoid_potential_lamb_1}, $\lambda$ is the ellipsoidal coordinate that grows perpendicular to the ellipsoid with principal radii $(c,b,a)$ along $(x,y,z)$ respectively. As shorthand for the different terms in Eq.~\ref{eq:ellipsoid_potential_lamb_1} I introduce the definitions:
	\begin{align}
		C(t) &= \frac{w}{2-\alpha_0}, \label{eq:ell_int_C} \\
		\alpha_0 &= \alpha(0, a,b,c;\ t),\text{ and} \label{eq:ell_int_alpha0}\\
		\alpha(\lambda, a,b,c;\ t) &= abc \int_\lambda^\infty \frac{d\lambda'}{(a^2 + \lambda')^{3/2}(b^2+\lambda')^{1/2}(c^2+\lambda')^{1/2}} \label{eq:ell_int_alpha}
	\end{align}
	
	It should be noted that $x,y,z$ and $\lambda$ are defined with respect to the moving ellipsoid center. It therefore makes sense to consider the equivalent problem of the ellipsoid in an unsteady free-stream flow with changing velocity $-w(t)\ \zhat.$
	
	A point in ellipsoidal coordinates is defined by the variables $(\lambda, \mu, \nu)$ such that surfaces of constant $\lambda$ are ellipsoids offset from the primary ellipsoid with principal radii $(c,b,a)$ in the $(x,y,z)$ directions respectively. To convert a point $(x,y,z)$ from cartesian coordinates into ellipsoidal coordinates, the variables $\lambda, \mu, \nu$ can be found as the solutions of the third-order polynomial in $k$ defined by:
	
	$$ \frac{z^2}{a^2 + k} + \frac{y^2}{b^2 + k} + \frac{x^2}{c^2 + k} - 1 = 0. $$
	
	Canonically, the solutions for $k$ are sorted so that $\lambda > \mu > \nu.$ With this sorting, $\lambda$ represents the spatial variable that increases perpendicular to the ellipsoid surface; surfaces of constant $\lambda$ take the form of ellipsoids with scaled principle radii. The surface $\lambda=0$ lays coincident with the surface of the ellipsoid $(c,b,a)$. 
	
	For the case of an oblate ellipsoid of revolution (about the $\zhat$ axis) with $a<b,$ $b=c$ and $x^2 + y^2 = r^2,$ this equation reduces to: 
	
	\begin{equation} \label{eq:ell_coord_transformation}
		\frac{z^2}{a^2 + k} + \frac{r^2}{b^2 + k} - 1 = 0.
	\end{equation}
	
	Since we are only interested in the pressure on the surface ($\lambda=0$) of such an oblate, axisymmetric ellipsoid translating along $\zhat$, we can also use these assumptions to simplify Eq.~\ref{eq:ellipsoid_potential_lamb_1}:
			
	\begin{equation}\label{eq:ellipsoid_potential_lamb_simp}
		\phi(x,y,z,t) = - z \cdot C(w, a, b; t) \cdot \alpha(\lambda=0, a,b;\,z,r,t).
	\end{equation}
	
	\noindent Furthermore, this simplification admits the following closed-form solution to the integral: 
	
	\begin{align}
	\alpha(&\lambda, a,b;\ t) \nonumber \\ 
	&= a b^2 \int_\lambda^\infty \frac{d\lambda'}{(a^2 + \lambda')^{3/2} (b^2 + \lambda)} \label{eq:alpha_integral}\\
	& = ab^2 \left[ -\frac{\pi}{(b^2-a^2)^{3/2}} + \frac{2}{(b^2-a^2)(a^2 + \lambda)^{1/2}} + \frac{2 \tan^{-1}\left(\sqrt{\frac{a^2+\lambda}{b^2-a^2}}\right)}{(b^2-a^2)^{3/2}} \right]. \label{eq:alpha_integral_soln}
	\end{align}
	
	The integral $\alpha$ relates to the Bessel added mass $k_z,$ for the ellipsoid moving in the $z$-direction, according to \cite{Lamb:1945tl}:
	\begin{equation} \label{eq:added_mass_ellipsoid}
		k_z = \frac{\alpha_0}{2-\alpha_0}
	\end{equation}
	Using this relationship, it can be shown that the closed form solution Eq.~\ref{eq:alpha_integral_soln} approaches the expected limits as $a\to0$ and $a\to b$, corresponding to a flat plate and a sphere respectively.

	From these simplifications, it is possible to explicitly calculate the pressure on the surface of the vortex bubble using only the measured motion and growth of the ellipsoid bounding the vortex bubble. To calculate the spatial derivative required by the equation (Eq.~\ref{eq:bernoulli}) I observe that the only spatially dependent terms in Eq.~\ref{eq:ellipsoid_potential_lamb_simp} are $C$ and $\alpha$ so that:

	$$ \mathbf{\nabla} \phi(t) \Big|_{S_{vb}} = 
	-\zhat\cdot C(w, a, b; t)\cdot \alpha_0(a,b,t) 
	- z\cdot C(w, a, b; t) \cdot \frac{\partial \alpha}{\partial \lambda}\Big|_{\lambda=0}\cdot \left( \zhat \frac{\partial\lambda}{\partial z}\Big|_{\lambda=0} + \rhat \frac{\partial\lambda}{\partial r} \Big|_{\lambda=0}\right).$$

	\noindent The derivative $\partial \alpha / \partial \lambda$ can be evaluated by applying the fundamental theorem of calculus to Eq.~\ref{eq:alpha_integral}. The spatial derivatives of $\lambda$ can then be calculated by implicitly differentiating Eq.~\ref{eq:ell_coord_transformation} with $k \equiv \lambda.$	To calculate the time derivative of the velocity potential on the surface of the ellipsoid, the chain rule can be applied to $\phi:$

	$$ \frac{\partial \phi}{\partial t} \Big|_{S_{vb}} = -z\cdot\alpha(0,a,b; t)\cdot \left( \frac{\partial C}{\partial w} \dot{u}_{vb} + \frac{\partial C}{\partial a} \dot{a} + \frac{\partial C}{\partial b} \dot{b} \right)
	- z\cdot C(w,a,b; t) \cdot \left( \frac{\partial \alpha}{\partial a} \dot{a} + \frac{\partial \alpha}{\partial b} \dot{b} \right)\Big|_{\lambda=0}. $$

	\noindent Here, derivatives with respect to $w,a,$ and $b$ can be evaluated using Eqs.~\ref{eq:ell_int_C},\ref{eq:ell_int_alpha0}, and \ref{eq:alpha_integral_soln}.
	
	Once these two derivatives are calculated, they can be inserted into the Bernoulli equation to estimate the pressure on the leading surface of the vortex bubble.
	
	\section{Derivation of Induced Velocity Scaling}
	
	To describe the coupling of two simultaneous jets, we propose that the over-pressure at each nozzle is modified by the ``induced'' pressure, $p_{ind}\sim \rho u_{ind}^2,$ associated with the other jet's developing vortex ring. Here, we derive our scaling for the induced velocity, $u_{ind}\sim r^{-3}.$ 
	
	\begin{figure*}[h]
		\centering
		\includegraphics[width=0.4\textwidth]{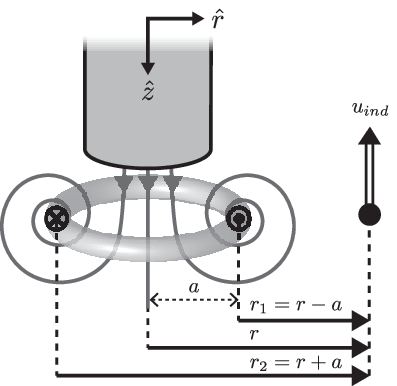}
		\caption{\label{fig:supp:InductionGeom} Geometry for induced velocity derivation.}
	\end{figure*}
	
	In 3D, a toroidal vortex ring with circulation $\Gamma$ and radius $a$ is described exactly by the axisymmetric stream function (See \citet{Lamb:1945tl}, Art.~161):
	
	\begin{align}
	\psi(r, z) &= - \frac{\Gamma}{2\pi}(r_1 + r_2)\left[K(\lambda) - E(\lambda) \right]
	\end{align}
	\noindent where
	\begin{align*}
	\lambda = \frac{r_2 - r_1}{r_2 + r_1}, \qquad
	r_1^2 = z^2 + (r-a)^2, \quad \text{and} \quad
	r_2^2 = z^2 + (r+a)^2
	\end{align*}
	\noindent As shown in Fig.~\ref{fig:supp:InductionGeom}, $r_1$ is the closest distance of the vortex ring to the field point $(r,z)$, and $r_2$ is the farthest distance. $K$ and $E$ are complete elliptic integrals of the first and second kind, respectively.
	
	Since the two vortex rings form simultaneously, we expect that their mutual effects are dominant on the plane $z=0$. In this case, the induced velocity is antiparallel to $\zhat$, $r_1=r-a$, $r_2=r+a$, and $\lambda = a/r = D/(2r).$ Differentiating the stream function yields a result for the velocity induced by a developing vortex ring:

	\begin{align}\label{eq:supp:induced_vel}
	\mathbf{u}_{ind}(r, z=0) &= \frac{1}{r} \frac{\partial \psi}{\partial r} \zhat = - \zhat \left(\frac{2 \Gamma}{\pi D}\right) \lambda \left[K\left(\lambda\right) - \frac{2-\lambda}{2(1-\lambda)}E\left(\lambda \right) \right] .
	\end{align}
	
 The complete elliptic integrals can be expanded in a power series about $\lambda=0$ \cite{Friedman:1971cl}:
	
	\begin{align*}
	K(\lambda) = \frac{\pi}{2} \left[ 1 + \frac{1}{4} \lambda^2 + O(\lambda^4)\right] \qquad
	E(\lambda) = \frac{\pi}{2} \left[ 1 - \frac{1}{4} \lambda^2 - O(\lambda^4) \right].
	\end{align*}

	Inserting these expressions into Eq.~\ref{eq:supp:induced_vel}, expanding $u_{ind},$ and substituting $\lambda = D/(2r),$ the induced velocity can be written to leading order as
	
	\begin{align}
		\mathbf{u}_{ind} (r) \approx (-\zhat) \left(\frac{\Gamma}{2 D} \right) \left(\frac{D}{r}\right)^{3}.
	\end{align}

	This leading order scaling is valid as long as $\lambda \ll 1$. As $\lambda \to 1,$ $u_{ind}$ diverges to $+\infty$ and higher order terms dominate. In terms of $r/D,$ the higher order terms should be negligible until $r/D \lesssim 1.$ In the regime of our experimental data, $r/D \gtrsim 1.7,$ the next term, of $O(r^{-4})$, is 3 times smaller than the leading order term. Hence we describe the physics of the two-jet coupling using the leading-order approximation, $u_{ind}\sim r^{-3}$.
		
	\section{Additional Measurements}
	
	\subsection{Wake Kinematics}
	In our analysis, we asserted that the $z$-position and volume of the control volume grow linearly with time, so that higher order derivatives such as $\dot{u}_{cm}$ and $\ddot{V}_{cv}$ could be ignored. Here we provide typical data to support this claim. As shown in Fig.~\ref{fig:supp:CVMeasurements}, both position and volume are roughly linear across the time range measured. Small sinusoidal oscillations arise in the plots from a shear instability along the jet (visible in the videos). By fitting these quantities to a line, the oscillations are suppressed. While the wake position and volume are roughly linear over the experiment, the vortex bubble shape is not. The major and minor radius of the bounding ellipsoid are tracked during the experiments, and their values are used directly to compute the time-dependent pressure on the leading edge of the wake.
	
	\subsection{Thrust}
\begin{figure*}[b]
		\centering
		\includegraphics[width=\textwidth]{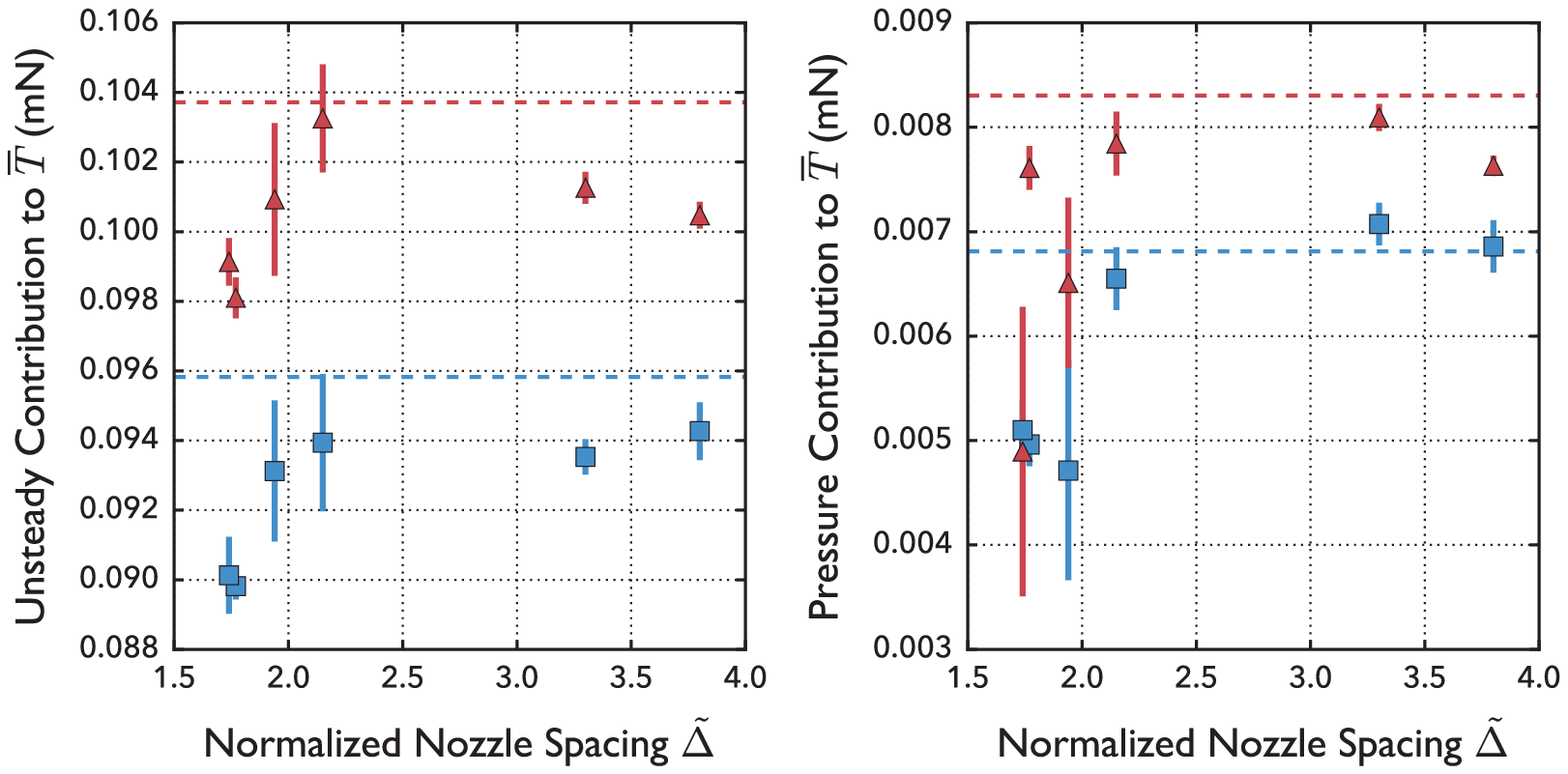}
		\caption{\label{fig:supp:ThrustContrib} Contribution of the two different terms, unsteady and surface pressure, to the thrust measurement. Points represent the average of 5 experiments for each value of $\tDelta.$ The dotted lines represent the average of $5$ experiments for the single nozzle experiments - nozzle 2 is indicated by red and nozzle 1 is indicated by blue. The unsteady contribution is typically 10 times stronger than the pressure contribution.}
	\end{figure*}

\begin{figure*}
		\centering
		\includegraphics[width=0.65\textwidth]{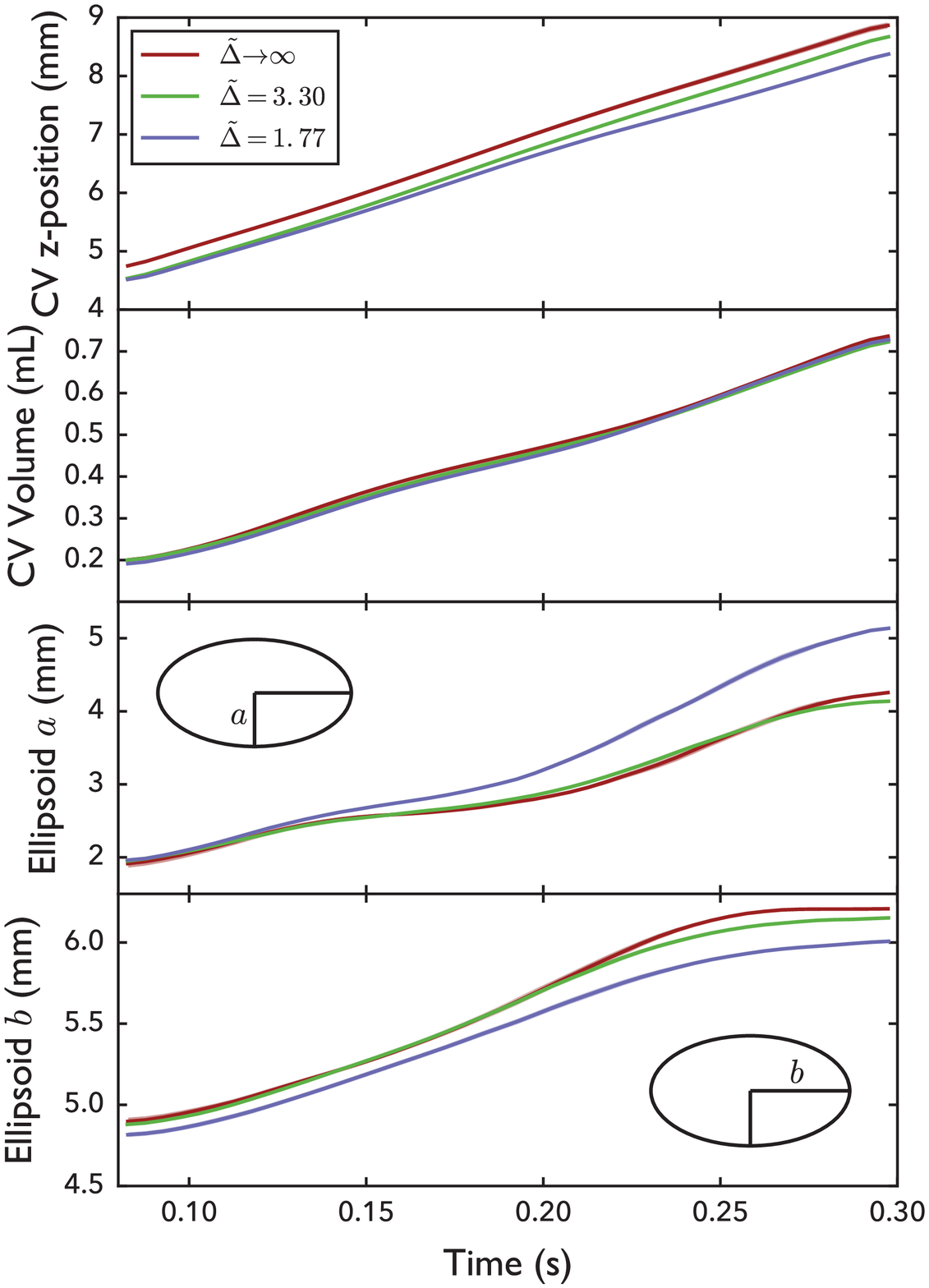}
		\caption{\label{fig:supp:CVMeasurements} Results of CV tracking for select experiments. Solid lines represent the median curve from 5 experiments, and the shaded regions represent one standard deviation. As indicated by the small error bars, these experiments produced highly repeatable results.}
	\end{figure*}

	In the main text, we showed only normalized thrust measurements, $T(\tDelta)/T_\infty,$ but claim that the contribution from the pressure integral is an order of magnitude smaller than the contribution from unsteadiness in the wake. In Fig.~\ref{fig:supp:ThrustContrib}, we provide the data to support this claim.
	
	\bibliography{bibliography}